# A Hardware-Efficient Synchronization in L-DACS1 for Aeronautical Communications

Thinh Hung Pham, *Member, IEEE*, Vinod A. Prasad, *Senior Member, IEEE*, and A. S. Madhukumar, *Senior Member, IEEE*

*Abstract*—*L*-band digital aeronautical communication system type-1 (L-DACS1) is an emerging standard that aims at enhancing air traffic management by transitioning the traditional analog aeronautical communication systems to the superior and highly efficient digital domain. L-DACS1 employs modern and efficient orthogonal frequency-division multiplexing (OFDM) modulation technique to achieve more efficient and higher data rate in comparison to the existing aeronautical communication systems. However, the performance of OFDM systems is very sensitive to synchronization errors such as symbol timing offset (STO) and carrier frequency offset (CFO). STO and CFO estimations are extremely important for maintaining orthogonality among the subcarriers for the retrieval of information. This paper proposes a novel efficient hardware synchronizer for L-DACS1 systems that offers robust performance at low power and low hardware resource usage. Monte Carlo simulations show that the proposed synchronization algorithm provides accurate STO estimation as well as fractional CFO estimation. Implementation of the proposed synchronizer on a widely used field-programmable gate array (FPGA) (Xilinx xc7z020clg484-1) results in a very low hardware usage which consumed 6.5%, 3.7%, and 6.4% of the total number of lookup tables, flip-flops, and digital signal processing blocks, respectively. The dynamic power of the proposed synchronizer is below 1 mW.

## I. INTRODUCTION

OVER the past two decades, the air transport industry has experienced continuous growth and the demand for passenger air traffic is forecast to double the current level by about 2025 [1]. The current air transportation systems will not be able to cope with this growth, e.g., already very high-frequency communication capacity is expected to saturate in Europe by 2020–2025 [2]. Meeting these growing demands require efficient air-to-ground communication systems providing various data of airplanes in real time. *L*-band digital aeronautical communication system (L-DACS) is being proposed as a solution that can coexist with legacy *L*-band systems and aims to explore digital radio techniques to enable efficient communication for next-generation global ATM systems [3]. There are two specifications that are being reviewed for L-DACS: type-1 (L-DACS1) and type-2 (L-DACS2). The type-1 specification defined by EUROCONTROL [4] is the most promising and mature candidate for final selection. L-DACS1 is intended to be operated in the frequency range of 960-1164 MHz that is mainly utilized by aeronautical navigation aids [e.g., the distance measuring equipment (DME) and the military tactical air navigation (TACAN)]. It might be difficult to allocate new spectrum due to the many existing *L*-band systems. The inlay approach where L-DACS1 is operated in the spectral gaps between two adjacent DME or TACAN channels is an attractive solution. This approach does not require any new spectrum allocation and the existing allocated spectrum of all other *L*-band systems can be retained. However, the inlay concept poses special interference issue. The L-DACS1 receiver has to be robust against interference from the other *L*-band systems. L-DACS1 employs orthogonal frequency division multiplexing (OFDM) modulation that has widely been employed in various high bit rate wireless transmission systems (e.g., WiMAX and WiFi). Therefore, L-DACS1 is able to achieve more efficient and higher data rate in comparison to the existing systems for ATM. OFDM is a modern and effective multicarrier modulation technique with its advantages of combating impulsive noise, robustness to multipath effects, and spectral efficiency. However, OFDM performance is sensitive to receiver synchronization. Carrier frequency offset causes intercarrier interference and errors in timing synchronization can lead to intersymbol interference (ISI) [5]. Therefore, synchronization is critical to the performance of OFDM systems such as L-DACS1.

Many techniques have been proposed for effective OFDM synchronization in the literature that include both data-aided schemes [6]–[8] and blind schemes [9], [10]. The latter schemes use inherent structure of the OFDM symbol (i.e., cyclic prefix). However, these methods usually require a large number of OFDM symbols leading to long delay and large computation cost to achieve satisfactory performance. In contrast, the former schemes employ training symbols (i.e., preamble) for either the autocorrelation of received preamble or the cross correlation between the copy of transmitted preamble and received symbol at the receiver. Autocorrelation-based methods for timing synchronization [8] rely on the repetitions of preamble that offer robustness to large carrier frequency offset (CFO) and multipath channel with low computational cost. These methods are suitable for coarse symbol timing offset (STO) and CFO estimation. In the absence of CFO, cross-correlation-based methods achieve an excellent timing synchronization performance [6]. However, the performance of these methods degrades significantly due to the presence of large CFO. This limits their use to fine timing estimation schemes [7], [11], [12], which use autocorrelation for coarse timing and cross correlation for fine timing estimation. L-DACS1 employs the preamble that has a similar structure to that of WiMAX but shorter length. In addition, L-DACS1 transmission is operated in *L*-band aeronautical channels which suffer from large interference and large Doppler shifts. Therefore, the synchronization for L-DACS1 poses more challenge to achieve a good performance in the *L*-band aeronautical channels. However, synchronization of L-DACS1 is hardly addressed in the literature. To the best of the authors' knowledge, there is only one synchronization method that is presented for an optimized L-DACS1 receiver [13]. This method is based on blind schemes leading to requirements of long synchronization time and large computation cost, and it just tolerates a small CFO which is less than one subcarrier spacing. Recently, we proposed an efficient data-aided synchronization method for L-DACS1 in [14]. This method is robust against large CFO and achieves accurate STO estimation as well as fractional CFO estimation in a range. However, no specific implementation for performing effectively the synchronization on hardware was dealt in [14].

In this paper, we extend the work in [14] to propose an efficient hardware implementation for an accurate L-DACS1 synchronizer. The implemented synchronizer is able to complete the computation of both frequency offset estimation and timing synchronization within the duration of the L-DACS1 preamble. In an aerospace environment, integrated circuits operating in an environment with radiation are sensitive to transient faults caused by the interaction of ionizing particles with silicon. The high amount of resources and flexibility of FPGA are very attractive for aerospace applications that require a high level of reliability [15]. Different fault tolerance techniques can be applied to FPGAs to provide a high reliability level. Some techniques are based on spatial redundancy. The spatial redundancy is based on the replication of $n$ times the original module. For example, triple modular redundancy, a most common case of $n$-modular redundancy, implements three identical modules to increase fault tolerance. This leads to significant increase of hardware and power consumption. The proposed method introduces an efficient synchronizer implementation with the proposed low power, low hardware usage architecture. This allows a promising realization of low-power OFDM-based systems on FPGA platform [16] for L-DACS1 communications.

The main contributions of this paper are as follows.
1) An accurate synchronization algorithm for L-DACS1 system that can yield excellent synchronization performance in the face of a large CFO.
2) An efficient hardware architecture for the L-DACS1 synchronizer which effectively performs the proposed algorithm with low hardware cost and low-power consumption. This is the first hardware implementation for the L-DACS1 synchronizer.
3) An enhanced circuitry which is based on an optimized direct form targeting for multiplierless correlation that results in a significant reduction on hardware cost in comparison to the state-of-the-art (SoA) multiplierless-transpose-form correlator.
4) An evaluation of estimation accuracy in case of optimized precision for metric calculation in L-DAC1 synchronizer.
5) A comparison to existing implementation approach for L-DAC1 synchronizer in terms of hardware usage and power consumption, and a comparison of resource utilization for computing the metrics used in the synchronizer.

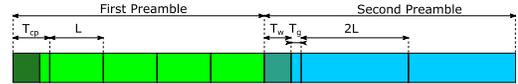

Fig. 1. Structure of the LDACS1 preamble symbols.

This paper is organized as follows. Section II discusses OFDM synchronization and outlines the proposed method, which is then simulated and evaluated in Section III. The hardware implementation is presented and discussed in Section IV before Section V concludes this paper.

## II. LDACS1 SYNCHRONIZATION

OFDM synchronization consists of STO estimation and CFO estimation. STO estimation is to find the first sample of each OFDM symbol to retrieve a complete OFDM symbol for demodulation. CFO estimation determines the frequency mismatch between transmitted samples at transmitter and received samples at receiver. Synchronization can be performed at receiver based on the special symbols, known as preamble, that are sent at the beginning of each physical frame. L-DACS1 is specified with two preamble symbols. The first symbol has a time domain waveform consisting of four identical parts of length $L$, whereas the second symbol is formed with two identical parts of length $2L$. L-DACS1 is specified with $L = 16 * N_{ov}$ where $N_{ov}$ is the oversampling factor in the receiver which is typically four [4]. The preamble of L-DACS1 in time domain after adding the cyclic prefix ($T_{cp}$) and applying windowing ($T_w$) is depicted in Fig. 1.

A signal is transmitted through a frequency-selective channel and corrupted by a zero-mean complex white Gaussian noise $\eta_n$. At the receiver, the signal is received and down converted to baseband for demodulation. The samples of received signal can be expressed as follows:

$$r_n = e^{j(2\pi \epsilon n/N)} \sum_{l=0}^{C-1} h_l x_{n-l} + \eta_n \quad (1)$$

where $x_n$ and $r_n$ denote transmitted and received samples, respectively. $h_l$ is the baseband equivalent discrete-time channel impulse response of length $C$. $\epsilon$ denotes the normalized carrier frequency offset between transmitter and receiver.

### A. Proposed Timing Metrics

We propose new timing metrics that take advantage of L-DACS1 preamble structure and energy correlation.

We define two autocorrelation metrics based on the periodic parts of the first preamble symbol as follows:

$$AC1(n) = \sum_{m=0}^{2L-1} c1(m,n)$$

$$AC2(n) = \sum_{m=0}^{2L-1} c2(m,n)$$

$$c1(m,n) = r^*_{n-m} r_{n-m-L}$$

$$c2(m,n) = r^*_{n-m} r_{n-m-2L} \quad (2)$$

where * denotes the complex conjugation. The first autocorrelation metric searches for quarter repetition, whereas the second finds half repetition of the first preamble symbol. These metrics are employed not only for frame detection but also for accurate CFO estimation which is discussed in Section II-B.

An energy metric, ENE is proposed which measures the received symbol energy in a duration of length $2L$. This metric is used as a reference for detect an incoming preamble

$$\text{ENE}(n) = \sum_{m=0}^{2L-1} r^*_{n-m} * r_{n-m}. \quad (3)$$

To keep the exposition simple, assume an ideal channel with noise. Then, samples of the received preamble are expressed as follows:

$$r_n = e^{j(2\pi\epsilon n/N)} * p_n + \eta_n \quad (4)$$

where $p_n$ denotes the transmitted samples of the preamble symbols. The metrics in (2) is derived as follows:

$$AC1(n) = e^{-j\phi_1} \sum_{m=0}^{2L-1} p^2_{n-m} + \eta_1(n)$$

$$\eta_1(n) = \sum_{m=0}^{2L-1} \eta^*_{n-m}\eta_{n-m-L} + p_{n-m}(\eta^*_{n-m} + \eta_{n-m-L}) \quad (5)$$

where $\eta_1(n)$ presents the noise part on the metric. $\phi_1 = 2\pi\epsilon(L/N)$ is the phase rotation caused by CFO. Similarly, $AC2(n)$ and $ENE(n)$ can be derived as follows:

$$AC2(n) = e^{-j\phi_2} \sum_{m=0}^{2L-1} p^2_{n-m} + \eta_2(n)$$

$$\text{ENE}(n) = \sum_{m=0}^{2L-1} p^2_{n-m} + \eta_e$$

where $\eta_2(n)$ and $\phi_2 = 2\pi\epsilon(2L/N)$ are the noise part and phase rotation part of the second autocorrelation metric. $\eta_e$ denotes the noise part of energy metric. If the noise part and phase rotations part are negligible, the autocorrelation metrics equal the energy metric when the first preamble symbol is received.

An energy correlation metric is presented for accurate STO estimation instead of using signal cross correlation as in conventional approach. The energy correlation is insensitive to phase rotation caused by CFO. Moreover, the metric is computed on real numbers leading to reduce computational requirement in comparison to the signal cross correlation. The periodic waveform of preamble symbols causes the peaks of cross correlations. Not only a peak occurs at the correct STO but also there are secondary peaks which are present at the repetition of waveform. The secondary peaks may cause incorrect STO estimations. The proposed energy correlation metric is expressed as follows:

$$\text{XCR}(n) = \sum_{m=0}^{D-1} |c2(m,n)| * a_m. \quad (6)$$

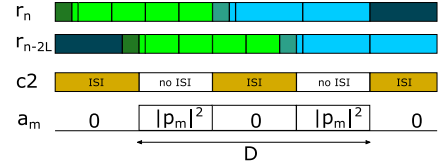

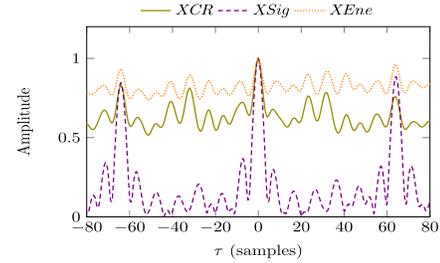

Fig. 2. Proposed correlation metrics on the L-DACS1 preamble.

Fig. 3. Cross-correlation metrics for STO estimation at SNR = 10 dB.

The proposed metric uses $|c2(m,n)|$ for energy correlation instead of instant energy $r^*_{n-m} * r_{n-m}$ like in [8]. $a_m$ is an energy vector of the transmitted preamble. $D$ is the length of the vector $a_m$ illustrated in Fig. 2. It should be noted that when preamble symbols presents at received signal, $|c2(m,n)|$ equals the instant energy. However, $|c2(m,n)|$ eliminates the periodic feature of the preamble as shown in Fig. 2 because ISI occurs when received signal multiplies with its delayed version. Therefore, the proposed metric can reduce the secondary peaks leading to improved STO estimation accuracy.

Fig. 3 shows the normalized values of the proposed metric (*XCR*) in comparison to signal cross correlation (*XSig*) and instant energy correlation (*XEne*). As can be seen, $\tau = 0$ corresponds to the correct STO where the largest peak occurs. The secondary peaks of the proposed metric located at $-64$, $64$ (i.e., $-L$ and $L$) is reduced compared to *XSig* and *XEne*.

### B. Proposed Synchronization Flow

The proposed method calculates the aforementioned timing metrics for jointly estimating STO and CFO in an efficient synchronization flow as follows. First, incoming preamble symbols are detected by a comparison operator between the autocorrelation metrics with the signal energy metric. Then,

STO estimation is performed using the proposed energy correlation metric and CFO is estimated by using the autocorrelation metrics. The preamble detection is based on the first preamble symbol. When the first symbol presents in received signal, the values of autocorrelation metrics are increased and the following condition is met:

$$AC(n) > ENE(n) \tag{7}$$

where $AC(n) = |AC1(n)| + |AC2(n)|$. To increase stability, the first preamble symbol is detected when condition (7) is met for $m$ consecutive samples (with $m = 8 * N_{ov}$ used throughout this paper). After the first preamble symbol is detected, the synchronization jointly performs STO and CFO estimations.

*1) STO Estimation:* According to (6), when the preamble is received and $c2(m,n)$ matches with $a_m$, XCR will get the largest peak. Therefore, STO can be determined by searching the largest peak of XCR metric. The STO estimation is performed as follows:

$$\hat{n} = \arg\max_{n \in \Delta} XCR(n) \tag{8}$$

where $\hat{n}$ denotes estimated STO. $\Delta$ is a searching window and empirically chosen to equal $56 * N_{ov}$ based on Monte Carlo simulations. If this number is smaller than 56, the searching window may not include the right peak of XCR metric leading to reduce the estimation accuracy. When this number is larger than 64 (i.e., the length of the periodic parts in the first preamble), the searching window includes the secondary peaks. This results in increasing the possibility of wrong peaks detection.

*2) CFO Estimation:* According to (5), CFO can be estimated using

$$\begin{aligned}\hat{\epsilon} &= \frac{\phi_1}{2\pi}\frac{N}{L} - z_1\frac{N}{L} \\ &= \frac{\phi_2}{2\pi}\frac{N}{2L} - z_2\frac{N}{2L}\end{aligned} \tag{9}$$

where $\phi_1 = \angle AC1$ and $\phi_2 = \angle AC2$. $z1$ and $z2$ are an integer. The angle of *AC1* or *AC2* can be used to accurately estimate the CFO if $z1$ or $z2$, respectively, equals zero. Because of $-\pi < \phi_1$ and $\phi_2 < \pi$, the estimation using *AC1* is limited in $\pm 2$ subcarrier spacing, while using *AC2* the estimation range is $\pm 1$ subcarrier spacing, respectively. When CFO is larger than the estimation range, $z1$ and $z2$ are nonzero that represent integer CFO that can be estimated as in [17]. Integer CFO estimation is beyond the scope of this paper. CFO estimation using *AC1* has larger range than that using *AC2*. However, the estimation using *AC2* is more accurate compared to *AC1*. Moreover, using *AC2*, CFO estimation can be performed on both preamble symbol leading to improve estimation accuracy. The proposed method combines both metrics to achieve the CFO estimation with wide range and high accuracy. The proposed CFO estimation is expressed as follows:

$$\hat{\epsilon} = \begin{cases} \frac{\phi_2}{\pi}, & \text{if } -\pi/2 < \phi_1 < \pi/2 \\ \frac{\phi_2}{\pi} + 2, & \text{if } \pi/2 < \phi_1 \\ \frac{\phi_2}{\pi} - 2, & \text{otherwise.} \end{cases} \tag{10}$$

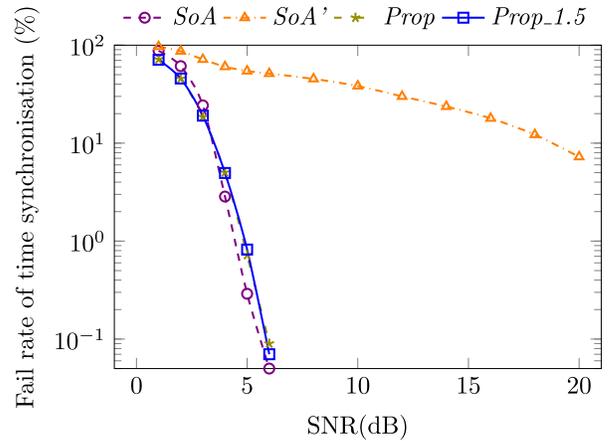

Fig. 4. Performance of time synchronization in AWGN channels with a frequency offset.

## III. SIMULATION

Monte Carlo simulations are performed to evaluate the proposed method for L-DACS1 systems. We investigate the synchronization performance in MATLAB under both AWGN channels and aeronautical propagation channels. 10 000 trials are simulated for each case. The accuracy of synchronization method is measured in terms of fail rate and mean square error (MSE) for STO and CFO estimations, respectively. Fail rate equals the fraction of unsuccessful timing estimations over the total trials (i.e., 10 000 times). MSE is calculated based on the error between CFO estimation and actual CFO.

### A. Performance in AWGN Channels

The performance of the proposed method is investigated in comparison to the SoA method presented in [13] in AWGN channel with the present of CFO. The comparison is presented in terms of the accuracy of both time synchronization and fractional CFO estimation. The performance of STO estimation is measured in terms of failure rate (%), and the accuracy of CFO estimation is evaluated in terms of MSE.

Fig. 4 shows the performance of timing synchronization in AWGN channels. *Prop* and *Prop_1.5* denote the proposed methods in cases of CFO absence and large CFO, respectively. The large CFO is set to equal 1.5 subcarrier spacing. The performance of the conventional method in [13] is denoted by *SoA* and *SoA'*. *SoA* is evaluated with the accuracy of coarse STO estimation. This means that a timing synchronization is successful if STO estimation is in cyclic prefix duration. But L-DACS1 requires the accuracy of STO estimation less than 1/11 cyclic prefix length [4]. *SoA'*, *Prop*, and *Prop_1.5* are measured for fine STO estimation in which a successful estimation requires an accuracy less than 1/11 cyclic prefix length. As can be seen in Fig. 4, the proposed method is robust against a large CFO. $Prop$ and $Prop\_1.5$ have almost identical fail rate. The proposed method achieves an excellent STO estimation with the presence of large CFO at SNRs above 5 dB. The proposed method outperforms the *SoA'* with the requirement of STO estimation accuracy less than 1/11 cyclic prefix length. The proposed method obtains better fail rate compared to the *SoA* at low SNRs (i.e., less than 3 dB). The *SoA* estimates for coarse STO, whereas the proposed method performs fine STO estimation. Hence, the *SoA* accuracy increases slightly faster than that of the proposed method, when SNR is larger. This results in a crossover at SNR of 4 dB. At SNRs larger than 6 dB, both the methods have excellent accuracy. In addition, a fine STO estimation is required following the *SoA* operation. This scheme requires further computation and long estimation time which is up to 0.25 s at high SNRs [13]. The proposed method can obtain the fine STO within the duration of preamble (i.e., 240 $\mu$s).

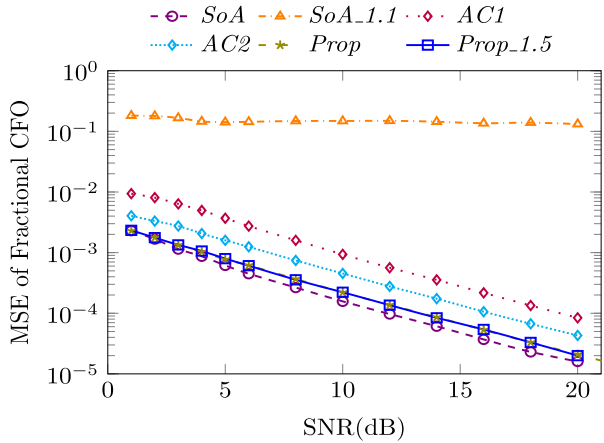

Fig. 5. Performance of frequency offset estimation in AWGN channels.

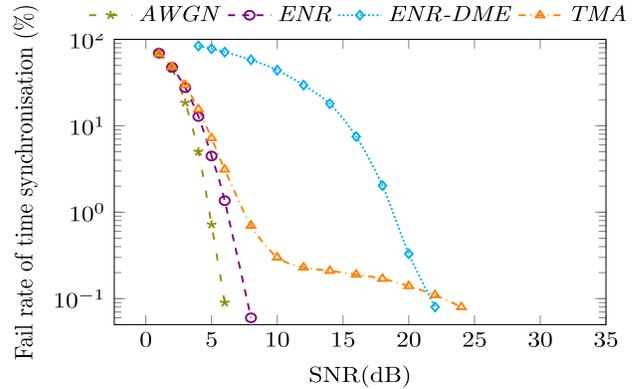

Fig. 6. Time synchronization performance in aeronautical channels.

Fig. 5 shows the accuracy of CFO estimation in AWGN channels. *SoA* and *SoA_1.1* denote the CFO estimation accuracy of the SoA method in cases of CFO absence and large CFO, respectively. The large CFO is set to equal 1.1 subcarrier spacing. *AC1* and *AC2* indicate the accuracy of the estimations using *AC1* and *AC2* metrics, respectively. These estimations are performed on the first preamble symbol. *Prop* and *Prop_1.5* present the estimation accuracy of the proposed method in cases of CFO absence and large CFO, respectively. The large CFO is set to equal 1.5 subcarrier spacing. As can be seen in Fig. 5, the CFO estimation using *AC2* metric is more accurate in comparison to *AC1* metric. The proposed method and SoA method have almost identical performance. Because these methods are performed on two preamble symbols instead of the first preamble symbol, *Prop* and *SoA* are more accurate compared to *AC2*. Moreover, *SoA_1.1* implies that the accuracy of the SoA method has significant degradation in case of large CFO. *Prop* and *Prop_1.5* are identical. This means that the proposed method maintains good performance even with CFO = 1.5 subcarrier spacing.

### B. Performance in Aeronautical Channels

In this section, the proposed synchronization method is further investigated in aeronautical channels. The channels are modelled in practical scenarios such as terminal manoeuvering (TMA) area and en-route (ENR) flight without/with DME interference. The simulation parameters for channel model are referenced from [13] and [18]. The channel models take into account many wireless channel effects including delay spread, Doppler spread, phase noise, and channel interference. The ENR channel was modeled with a strong line-of-sight path and reflected paths with a delay of $\tau 1 = 0.3$ $\mu$s and $\tau 2 = 15$ $\mu$s. The maximal velocity of airplane is assumed 1360 km/h, corresponding to a maximal Doppler shift of 1250 Hz. The TMA channel was modelled with maximum path delay of 10 $\mu$s, Rician factor of 10 dB, and maximal Doppler shift of 624 Hz. A realistic model presented in [19] is used to simulate DME interference for the area around Paris as this is the area with the highest density of DME ground stations in Europe. For this DME interference model, there are three DME interference sources with pulse rate of 3600 pulse pairs per second for each. The first DME channel locates at $-0.5$ MHz offset to L-DACS1 centre frequency. This channel causes interference of $-67.9$ dBm at L-DACS1 Rx input. The other two channels are at $+0.5$-MHz offset with interference of $-74$ and $-90.3$ dBm.

Figs. 6 and 7, respectively, depict the timing synchronization and CFO estimation performance of the proposed method in the mentioned scenarios of aeronautical channels in comparison to its performance in AWGN channels. *ENR* denotes the synchronization performance in an ENR channel without DME interference, while *ENR-DME* implies the case in an ENR channel with DME interference. *TMA* denotes the synchronization performance in TMA scenario. As can be seen in Figs. 6 and 7, in case of ENR channels without DME interference, the accuracy of STO and CFO estimations has a slight decrease compared to the case in AWGN channels. The proposed method can achieve excellent accuracy in the ENR channels at SNRs above 8 dB. However, when DME interference presents in ENR channels, the performance of the proposed method has considerable degradation. The accuracy of STO estimation and CFO estimation reduce about 10 and 4 dB, respectively. The timing synchronization can only achieve good performance at SNRs above 20 dB. In case of an TMA channel, the synchronization has almost similar performance compared to the case of AWGN and ENR channels at SNRs below 8 dB. The CFO accuracy of *TMA* saturates at SNRs above 10 dB while STO estimation of *TMA* slightly improve at SNRs above 10 dB and achieve an excellent accuracy at SNRs above 22 dB.

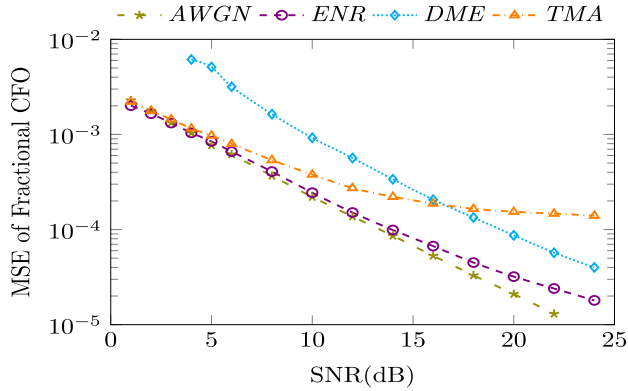

Fig. 7. Frequency offset estimation performance in aeronautical channels.

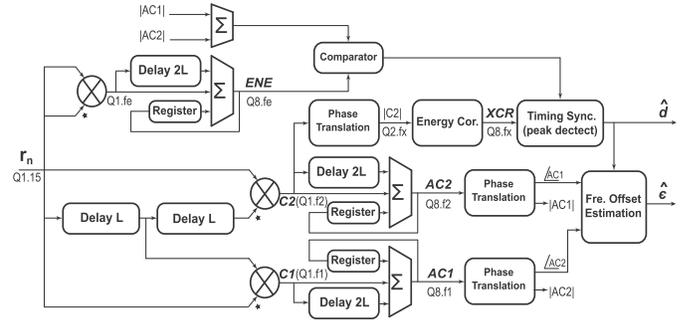

Fig. 8. Architecture for the implementation of the proposed synchronization method.

## IV. HARDWARE IMPLEMENTATION

This section discusses the implementation, and investigates the consumed hardware optimization, of the proposed method based on multiplierless techniques and word size precision. The target FPGA is a Xilinx Zynq-7000 xc7z020clg484-1 device, which is used in a development board named Zedboard, with Vivado 2016.1 used to evaluate both hardware resource and power consumption. The results illustrate the tradeoff between hardware consumption and the accuracy of the proposed method.

The implementation first presents the effective hardware architecture that performs the proposed synchronization method presented in Section II. Then, reduced precision of metrics values is carefully investigated to obtain the tradeoff between increasing hardware usage efficiency and reducing accuracy. To achieve further hardware reduction, a novel efficient circuitry for multiplierless correlation is proposed in Section IV-C. Finally, the implementation results were given with the comparison between alternative design optimizations and the comparison between computational modules in the proposed synchronizer.

### A. Implementation of Proposed Synchronizer

The architecture for our proposed method is shown in Fig. 8.

In implementation, the autocorrelation metrics in (2) and (3) are rewritten in efficient formulas to compute in a recursive manner as follows:

$$AC1(n) = AC1(n-1) + c1(0,n) - c1(2L, n-2L)$$
$$AC2(n) = AC2(n-1) + c2(0,n) - c2(2L, n-2L)$$
$$ENE(n) = ENE(n-1) + ee(n) - ee(n-2L) \quad (11)$$

where $ee(n) = r_n^* * r_n$. As can been seen in Fig. 8, these metrics are calculated by three-input adders. The *Delay 2L* blocks are employed to buffer the last *2L* values of *c1*, *c2*, *ene* for the corresponding metrics computation. *Registers* store the previous value of the metrics. A received buffer, including 2 blocks of *Delay L*, is shared for computing the operands *c1 and c2*. Three *Phase Translation* blocks are used to translate from rectangular form to polar form the values of *AC1, AC2*, and *c2* for CFO and STO estimations. We make use of the Xilinx CORDIC IP core to perform operation of the *Phase Translation* blocks. The *Energy Cor.* blocks calculates the *XCR* metric according to (6) based on multiplierless correlation. An enhanced architecture for multiplierless correlation proposed for this block is presented in Section IV-C. The *Fre. Offset Estimation* block is used to estimate the CFO according to (10) based on the angles of the *AC1 and AC2* metrics, received from the *Phase Translation* blocks. The *Timing Sync* block searches the peak of the *XCR* metric to estimate the STO when conditions checked by the *Comparator* block are met according to (7).

### B. Effect of Reduced Precision

In this section, word length optimization is investigated. The effect of reducing word length is considered to maintain the accuracy of estimation. We assume that the received samples are normalized and represented in the twos complement fixed point Q1.15 format. This means that the real and imaginary parts of sample have their values less than 1 and the values of each part are represented in total 16 bits with a sign bit and 15 fractional bits.

The values of instant autocorrelation *c1* and *c2* and instant energy *ee* are scaled to be less than 1 and let *f1, f2,* and *fe* be the number of bits representing the fractional part for computing *c1, c2,* and *ee*, respectively. This results in *c1, c2,* and *ee* be in the formats of *Q1.f1, Q1.f2,* and *Q1.fe*, respectively. Since the autocorrelation metrics are computed on the summation with the length of *2L* samples (i.e., 128 samples), the values of metrics are less than 128, needing just 8 bits for representation in twos complement to avoid the overflow of the summation. Consequently, the *AC1, AC2*, and *ENE* metrics are represented in *Q8.f1, Q8.f2*, and *Q8.fe*, respectively.

In addition, we considered the magnitude of *c2* (i.e., |c2|) be in format *Q2.fx*. The number of integer bits of |c2| equals 2 to avoid overflow due to the translation of complex number *c2*. The vector $a_m$ in (6) is normalized and quantized to 0.5 to employ multiplierless correlation. According to the L-DAC1 preamble, the vector $a_m$ has 132 nonzero coefficients. This results in the values of *XCR* be less than 256, needing just 8 bits positive number for representation to avoid the overflow. Therefore, the *XCR* metric is represented in *Q8.fx*.

The accuracy of CFO estimation depends on the autocorrelation metrics and the STO estimation requires the comparison between the autocorrelation metrics and the *ENE* metric.

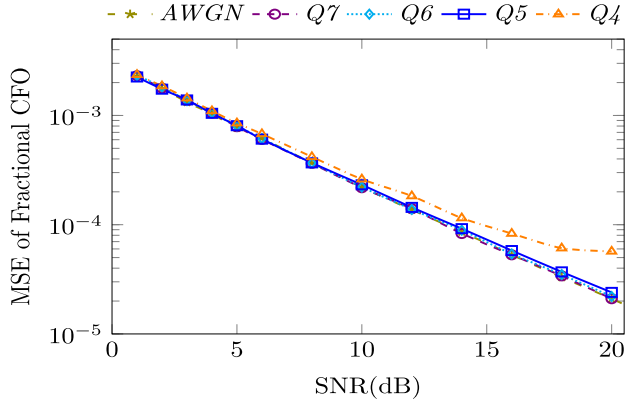

Fig. 9. Effect of reduced precision on CFO estimation performance.

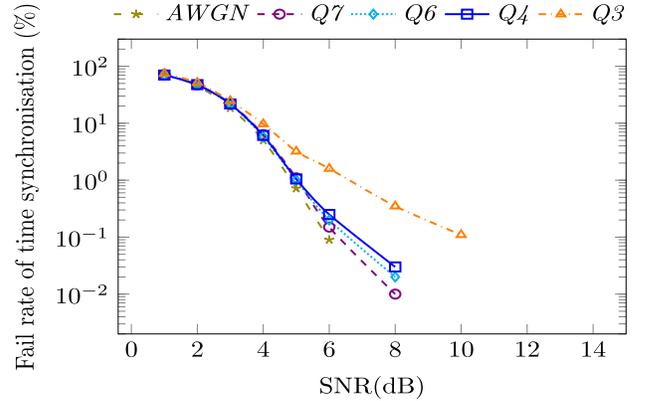

Fig. 10. Effect of reduced precision on STO estimation performance.

TABLE I
RESOURCES REQUIRED FOR COMPUTING AUTOCORRELATION METRIC ON FPGA WITH DIFFERENT WORD LENGTHS, $Q7.fa$

| $fa$ | 7 bits | | 6 bits | | 5 bits | | 4 bits | |
|---|---|---|---|---|---|---|---|---|
| | LUTs | FFs | LUTs | FFs | LUT | FFs | LUTs | FFs |
| AC1 | 76 | 203 | 69 | 201 | 62 | 199 | 55 | 197 |
| AC2 | 76 | 203 | 69 | 201 | 62 | 199 | 55 | 197 |
| ENE | 42 | 109 | 39 | 108 | 35 | 107 | 32 | 106 |
| Sum | 194 | 515 | 177 | 510 | 159 | 505 | 142 | 500 |

We consider these metrics has the same precision (i.e., same number of fractional bits). Let $fa = f1 = f2 = fe$ be a common number of fractional bits for the metrics. Fig. 9 illustrates the accuracy of the CFO estimation in reducing the number of fractional bits. *AWGN* denotes the results of metric calculation with full length computation (i.e., using 15 fractional bits), whereas *QN* present the performance of estimations in which the metrics are computed using $N$ fractional bits (i.e., $fa = N$ bits). Table I correspondingly reports the hardware usage in terms of the number of lookup tables (LUTs) and flip-flops (FFs) for computing the metrics with respect to different values of $fa$. By reducing $fa$ from 7 to 4 bits, the hardware cost is approximately linearly decreased, whereas the estimation accuracy are almost identical for the case of $fa$ equal 7, 6, and 5 bits. The accuracy has a considerable degradation in the case of $fa = 4$ bits. Consequently, we choose $fa = 5$ bits be an optimal trade-off point for reduce precision of the metric calculation.

The accuracy of STO estimation depends almost entirely on the *XCR* metric. This accuracy is investigated with regard to reducing the number of fractional bits of *XCR* (i.e., $fx$). Fig. 10 presents the accuracy of the STO estimation in reducing $fx$. *AWGN* denotes the results of *XCR* calculation with full length computation (i.e., using 15 fractional bits), whereas *QN* present the performance of estimations in which the *XCR* metric are computed using $N$ fractional bits (i.e., $fx = N$ bits). As can be seen in Fig. 10, reducing $fx$ from 7 to 5 bits leads to a negligible reduction of STO estimation accuracy. However, the degradation of the estimation becomes significant in the case of $fx = 3$ bits. Therefore, $fa = 4$ bits is chosen be an optimal trade-off point for reduce precision of the *XCR* calculation.

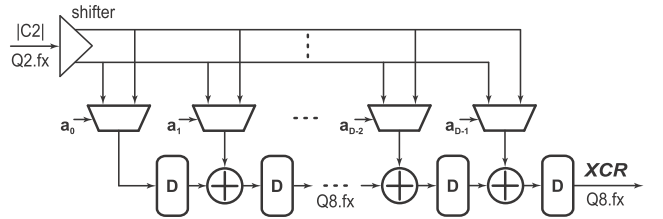

Fig. 11. Transpose circuitry of multiplierless correlation.

### C. Multiplierless Energy Correlators

The computation of correlation normally demands a large amount of resources in a synchronizer. To achieve further hardware reduction, we proposed a novel efficient circuitry relied on multiplierless correlation. The SoA multiplierless correlator is presented in [6] and [8] that is applied for IEEE 802.16. Fig. 11 illustrates the architecture of this correlator applied in the proposed synchronization. This architecture relies on a transpose form. The multiplication between input samples and constant coefficients, $a_m$, is realized by shift and add operator leading to eliminate the usage of actual multipliers. The transpose form is inherently pipelined and support multiple constant multiplications technique that results in significant saving of computation [20]. However, in case of multiplierless approach, this architecture has an intrinsic inefficiency in terms of word length of intermediate operands and operations. Indeed, as can be seen in Fig. 11, this architecture for the proposed method requires that the delay elements ($D$) and adders are implemented with the word length of $8 + fx$ bits including 8 integer bits to avoid overflow.

Considering the limitation of the transposed multiplierless correlator, this paper presents an enhanced architecture for multiplierless correlators relying on direct form. The architecture is designed based on the mathematical manipulations from (6) as follows:

$$\text{XCR}(n) = \sum_{m=0}^{D-1} |c2(m,n)|a0_m + 2^{-1}\sum_{m=0}^{D-1}|c2(m,n)|a1_m \quad (12)$$

where $a0_m$ and $a1_m \in \{0, 1\}$ represents the quantization levels of $a_m$. $a_m$ is quantized to 0.5 to employ multiplierless correlation. Therefore, $a_m$ can be rewritten as $a_m = a0_m + a1_m * 2^{-1}$.

Fig. 12 presents the proposed architecture for direct-form multiplierless correlation. As can be seen, the word length of delay element (*D*) is reduced to $2 + fx$ bits instead of $8 + fx$ bits. In addition, according to (12), *XCR* has two summation parts. Each part is computed using an adder network to minimize the word length requirement of adder. The network has *n* levels (*sn*) depended on number of elements (*Ne*) in the corresponding summation part, $n = \log_2 Ne$. At level *ii*, the adders require the word length of $2 + ii + fx$ bits instead of $8 + fx$ bits to avoid overflow.

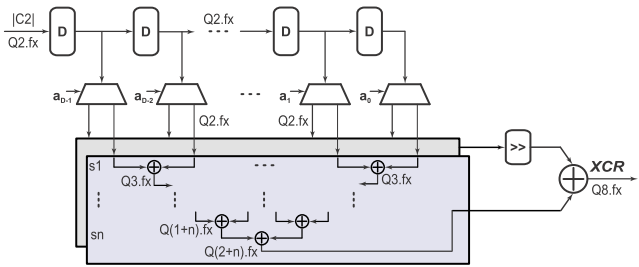

Fig. 12. Proposed circuitry for multiplierless correlation.

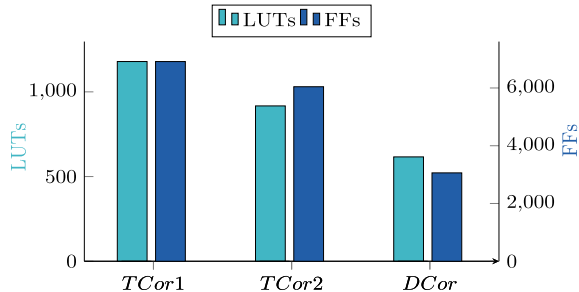

Fig. 13. Comparison of required resources for multiplierless correlation implementations.

Fig. 13 illustrates the hardware usage, in terms of number of LUTs and FFs, reduction for *XCR* calculation. *TCor1* and *TCor2* denote the hardware usage of the transposed correlator [8] with the reduced precision of $fx = 7$ bits and $fx = 4$ bits, respectively, whereas *DCor* presents the results of the proposed direct form architecture applied $fx = 4$ bits. As can be seen, *TCor2* has slight decrease in comparison to *TCor1*. Interestingly, *DCor* achieves a significant reduction, approximately *50% of TCor1*. Hardware usage reduction gained by enhancing architecture is more significant than that obtained by reducing the precision.

### D. Implementation Results

The preceding results are now used to define three alternative implementations for the proposed L-DAC1 synchronization to compare the hardware reduction obtained by employing the SoA circuitry [6] and that by applying the proposed circuitry. These alternatives are, namely, are as follows.

1) *Opt1:* A conventional instance of the proposed method applied the circuitry presented in [6] without optimized word length (i.e., both *fa* and *fx* set to 7).
2) *Opt2:* A hardware optimized instance which makes use of the circuitry presented in [6] with $fa = 5$ and $fx = 4$.
3) *Prop:* A proposed hardware efficient instance which is applied the proposed correlation circuity with $fa = 5$ and $fx = 4$.

TABLE II
TOTAL RESOURCES CONSUMED BY THREE REDUCED COMPLEXITY
INSTANCES OF THE PROPOSED METHOD

|      | LUTs | FFs  | DSPs | Qpwr   | Dpwr  | Fre   |
|------|------|------|------|--------|-------|-------|
| Opt1 | 4106 | 7846 | 14   | 119.95 | 0.326 | 11.36 |
| Opt2 | 3752 | 6933 | 14   | 119.95 | 0.306 | 12.05 |
| Prop | 3452 | 3950 | 14   | 119.95 | 0.277 | 11.79 |

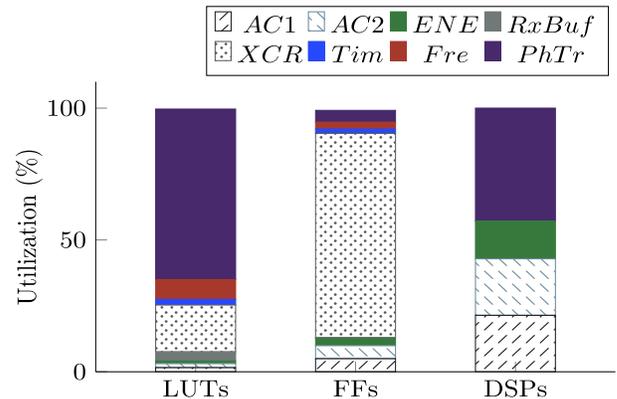

Fig. 14. Resource utilization of functional modules in hardware efficient synchronizer.

Table II reports the overall hardware resources required for these instances of the synchronizer in terms of the number of occupied LUTs, FFs, and digital signal processing (DSPs) slices. Dynamic (Dpwr) and quiescent (Qpwr) power consumptions are reported in mW. Maximum frequency (Fre) is reported in MHz. It should be noted that the accuracy of STO and CFO estimation of these instances can be seen by choosing the corresponding word length (i.e., *fa* and *fx*) from the plots of in Figs. 9 and 10, respectively. All instances have been simulated and reported in Section IV-B. From Table II, it is shown that the proposed alternative, *Prop*, achieves reductions of 16% and 50% of occupied LUTs and FFs in comparison to conventional implementation, *Opt1*. Comparing to the optimized instance applied the circuitry in [6], *Opt2*, the proposed synchronize has used hardware reductions of 8% and 43% of occupied LUTs and FFs. Overall, the hardware usage of the proposed synchronizer accounts for 6.5%, 3.7%, and 6.4% of the total number of LUTs, FFs, and DSP blocks, respectively, of the FPGA device (xc7z020clg484-1).

The power consumption is analyzed based on switching activities provided from postimplementation simulations with operating frequency of 2.5 MHz (i.e., $N_{ov} * 625$ KHz). The dynamic powers of three instances are less than *1 mW*. The maximum frequencies of three instances are also reported. The direct-form correlator typically results in considerable decrease of maximum frequency. However, the maximum frequency of the proposed synchronizer has a slight decrease compared to *Opt2*. Indeed, the required frequency for baseband processing in L-DAC1 systems is 2.5 MHz. This requirement is easily met by all tested implementations.

Fig. 14 details the comparison of resource utilization between the computations in the proposed hardware efficient synchronizer, *Prop*. The computation of *XCR* takes a large contribution of occupied resources (17.8% and 77.5% of LUTs and FFs, respectively). However, this computation avoids requirement of DSPs, whereas the computation of *AC1, AC2,* and *ENE* require 3, 3, and 2 DSP slices, respectively. The phase translation operation accounts for the largest percentage in terms of LUTs (64.7%) and DSPs (42.9%).

## V. Conclusion

L-DACS1 is being proposed as a solution that can coexist with legacy *L*-band systems and aims to explore OFDM-based radio techniques to enable high data rate communication for next-generation global ATM systems. This paper has presented and evaluated a hardware architecture of a novel synchronization method for L-DAC1. The proposed method is designed to achieve synchronization accuracy and to be much robust to large CFO than the SoA method. Moreover, the proposed method can obtain good synchronization in a short time within the duration of preamble. The accuracy of the proposed method is also demonstrated under several aeronautical channels. The implementation results show the proposed synchronizer employing a new efficient correlator achieves a reduced hardware usage and very low dynamic power consumption.

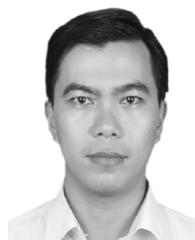

**Thinh Hung Pham** received the B.S. degree in electrical electronic engineering from the Ho Chi Minh City University of Technology, Ho Chi Minh City, Vietnam, in 2007, the M.Sc. degree in embedded systems engineering from the University of Leeds, Leeds, U.K., in 2010, and the Ph.D. degree from the Technische Universitat Munchen Program, Nanyang Technological University (NTU), Singapore, in 2015.

He was a Research Associate at the TUM CREATE Centre for Electromobility, Create Way, Singapore. From 2015 to 2016, he was a Lecturer at the Ho Chi Minh City University of Technology. He is currently a Research Fellow at the School of Computer Science and Engineering, NTU.

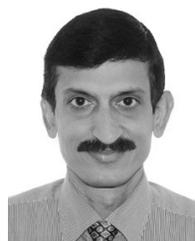

**Vinod A. Prasad** received the B.Tech. degree in instrumentation and control engineering from the University of Calicut, Malappuram, India, in 1993 and the M.Eng. (by research) and Ph.D. degrees from the School of Computer Engineering, Nanyang Technological University (NTU), Singapore, in 2000 and 2004 respectively.

He was an Automation Engineer at Kirloskar, Bangalore, India, at Tata Honeywell, Pune, India, and Shell, Singapore. From 2000 to 2002, he was a Lecturer at Singapore Polytechnic, Singapore. He joined the School of Computer Engineering, NTU, as a Lecturer in 2002, where he became an Assistant Professor in 2004, and was a Tenured Associate Professor from 2010 to 2017. He has been a Professor with the Department of Electrical Engineering, IIT Palakkad, Palakkad, India, since 2017.

Dr. Vinod was a recipient of the Nanyang Award for Excellence in Teaching in 2010, University's highest teaching recognition award.

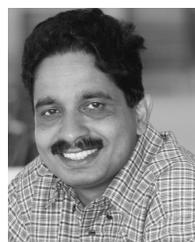

**A. S. Madhukumar** (SM'94) received the B.Tech. degree from the College of Engineering, Trivandrum, India, the M.Tech. degree from the Cochin University of Science and Technology, Kochi, India, and the Ph.D. degree from the Department of Computer Science and Engineering, IIT Madras, Chennai, India.

He was with the Center for Development of Advanced Computing, India, and the Institute for Infocomm Research, Singapore, where he was involved in communications and signal processing research. He is currently an Associate Professor at the School of Computer Science and Engineering, Nanyang Technological University, Singapore. He was involved in a number of funded research projects, organizing international conferences, and a permanent reviewer for many internationally reputed journals and conferences. He has authored over 200 referred international conference and journal papers.